# Geometry of Multihadron Production

JAMES D. BJORKEN

*Stanford Linear Accelerator Center*
*Stanford University, Stanford, California 94309*

ABSTRACT

This summary talk only reviews a small sample of topics featured at this symposium:

1. Introduction
2. The Geometry and Geography of Phase space
3. Space-Time Geometry and HBT
4. Multiplicities, Intermittency, Correlations
5. Disoriented Chiral Condensate
6. Deep Inelastic Scattering at HERA
7. Other Contributions

hep-ph/9411352

## 1. Introduction

At the time[1] of the first of these multiparticle conferences in Paris, 1970, only the first outlines of the present description of particle production at high energies were emergent. Descriptive tools, such as the rapidity variable, and basic concepts of limiting fragmentation (or Feynman scaling) had been in use for some time in the cosmic ray community, but only then were being recognized and developed further by the accelerator-based high energy community. And the understanding of the underlying strong interaction dynamics was of course primitive.

In the intervening years the change has been revolutionary. There is little question now that quantum chromodynamics (QCD) is the correct basic theory, and that the perturbative quark-gluon processes provide the correct dynamical picture for the origin of multiparticle production at sufficiently small distance scales. In addition, modern Monte Carlo computational methods have come of age and are an excellent match to the branching processes which dominate the perturbative calculations.

So it is appropriate to ask what it is that is left to do in this field. What are the most important areas left to study? To me the general answer is clear: it is those areas of QCD which are most *inaccessible* to theory which hold the most interest. Just as in QED, where the difficult-to-compute areas include all of condensed matter physics, chemistry, and biology, we may anticipate that the most novel dynamical features of QCD lie furthest away from the regions accessible to perturbation theory. At present, the identifiably difficult areas include the physics of the hadronization phase, everything to do with diffraction, and the high energy limit, at fixed transverse scale, of deep-inelastic structure functions (the BFKL limit). In addition, cosmic-ray anomalies at high energy, especially the claimed Centauro and "anti-Centauro" events exhibiting anomalous charge-fluctuations, hint at a possible connection with the spontaneously broken chiral phase of QCD, another poorly understood sector of the strong interaction.

In order to probe these difficult areas, one needs not only descriptive tools which are well-matched to the phenomena (these seem to exist in abundance) but also ways of isolating the crucial features of the underlying dynamics. I believe the geometry of the processes, both in momentum-space and in space-time, is essential for a solid interpretation of what is going on. In this talk this aspect will be emphasized. What follows is not really new. But it is a summary description, in my favorite language, of what I see as much of the common ground shared by the various QCD techniques now on the market.

## 2. The Geometry and Geography of Phase Space

At the time of the first multiparticle conference, the Feynman-fluid picture,[2] along with the mathematical machinery of generating functions and cumulants,[3] was being developed. The variables, $\eta$, $\phi$, $p_t$ became commonplace, along with the notion that only small $p_t$ and short-range correlations in rapidity were important. This implied that, in modern language, the mean density of hadrons in the lego plot approached a limit as $\ell n\, s$ became large, and consequently the "free energy" (the logarithm of the generating function $G(z)$ for multiplicity distributions) approached a thermodynamic limit[4]

$$\lim_{s \to \infty} \frac{\ell n\, G(z)}{\ell n\, s} = F(z) \ . \qquad (1)$$

While this description is not so bad at moderate energies, QCD modifies it in essential ways, because the quark/gluon jets carry large $p_t$ and because the spin-1 gluon exchanges can introduce long-range correlations in rapidity. An essential change is that the phase space becomes "extended", in fact fractal, and obtains an anomalous dimension. In addition the non-Abelian color degree of freedom introduces the specific rules of hadronization known as "angular ordering," "antenna rules," or "color-coherence."

## 2.1 Extended Phase-Space:

What follows is a variation[5] on a theme developed by the Lund group,[6] and specifically discussed in this meeting by Bo Andersson.[7] Consider $e^+e^- \to q\bar{q}g$ in a frame of reference collinear with $q$ and $\bar{q}$. The lego plot in this frame (see Fig. 1) has a gluon jet; a concentration of extra multiplicity is associated with that jet. To see it,

  $i$) Define the contents of the jet as what is inside a circle-of-radius-0.7 in the lego plot.
  $ii$) Introduce polar coordinates $\theta'$, $\phi'$ for the interior of the circle.
  $iii$) Introduce a new lego plot for the jet contents in terms of $\phi'$ and of
  $\eta' = -\ell n \tan \frac{\theta'}{2}$.

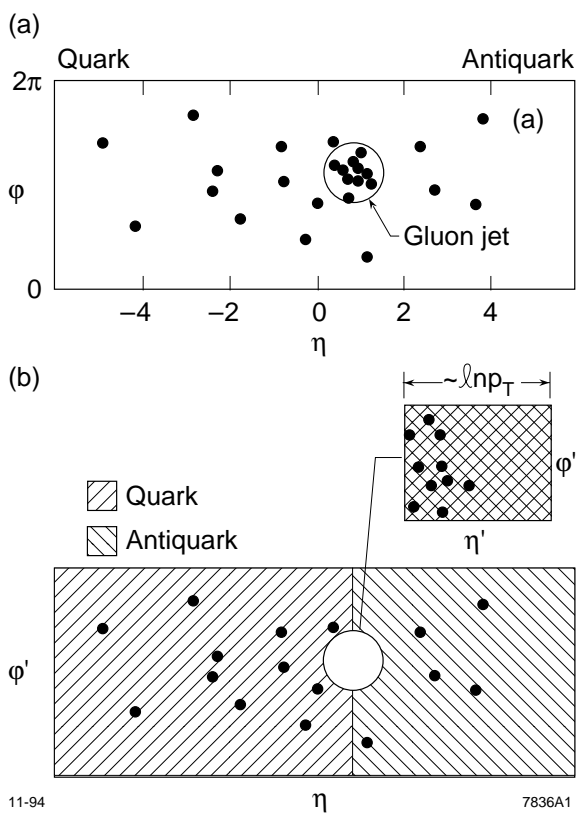

Figure 1. (a) Lego plot for the $q\bar{q}g$ final state in $e^+e^-$ annihilation, in a frame of reference for which the $q$ and $\bar{q}$ are collinear. (b) The same lego plot, with the extended phase space exhibited. Note the assignment of color to the phase-space regions affected by the $qq$ dipole (red) and and $g\bar{q}$ dipole (blue). The gluon phase-space extension is double-sided, with one color on each side.

The new plus old lego plot constitutes the *extended phase space*. Clearly if there are jets within the jets, the procedure can be iterated, and the fractal nature of the resulting phase-space should be clear. But once the jets—and minijets—are accounted for, we expect that the density of hadrons found in the extended phase space will be uniform, with limited correlations, and probably Poisson-like multiplicity fluctuations.[*] To my knowledge all models are consistent within this expectation.

*2.2 Color*

Introduction of the color degree of freedom leads to the well-known coherence effects.[8] To each region of the lego plot can be associated a color, that of the "string" or "color-dipole" which radiates quanta into that particular phase-space region. In particular, an important consequence is that in gluon jets the lego plot becomes "double-sided," because a gluon has two colors, not one, and (in the large $N_c$ approximation, which is quite good) each color radiates/fragments independently.[9] In the $e^+e^- \to q\bar{q}g$ example, again viewed in a reference frame where the $q$ and $\bar{q}$ jets are collinear (back-to-back), the color-dipoles associated with $q$ and $\bar{q}$ both radiate partons into the extended phase-space of the gluon. Here we put the products of $q$ on the front and that of $\bar{q}$ on the back. Consequently the gluon multiplicity at asymptotic $p_T$ is twice that for quarks (remember that $9/4 = 2 + \mathcal{O}(1/N_c^2)$).

All this is well known; the reason for mentioning it here however is the implication for multigluon production in hadron collisions. The population of gluons on the front of the lego plot, whatever their mutual correlation, remains *uncorrelated* with the population on the back. This feature has not been explored very much, but may well turn out to have nontrivial consequences for multiparticle/multijet correlation studies.

An amusing case is that of eclipsing jets. What is meant by this is a pair of gluon jets of similar $p_t$ on opposite sides of the lego plot but overlapping in $\eta$ and $\phi$. They will not merge into a single jet of twice the $p_T$ (and multiplicity $\approx \ell n\, 2p_t$) as occurs for the $e^+e^-$ case of a single-sided lego plot. Instead, they should independently fragment, leading to an asymptotic multiplicity $\approx 2\, \ell n\, p_t$, twice again as large as for a single merged gluon jet. It would be interesting to see, even if only in simulation, to what extent this distinction is observable.

---

[*] However, in *pp* collisions there will be at the very least broadening due to impact parameter fluctuations and/or constituent-quark substructure.

*2.3 Bottom Lines*

Our point of bringing up this rather familiar material is that these *general* features suffice to describe most of what is observed, irrespective of the further details of models. The important features include the following:

1. The nature of a QCD hard process is first controlled by what goes on at the highest $p_t$ (shortest distance) scale present in the event. This creates the basic architecture of the extended phase-space. Softer radiation decorates the basic architecture and creates the fractal-structure.

2. The amount of particle production is controlled by the total area of the extended phase space.

3. On the extended phase space one should find limited $p_t$ and only short range correlation structure, other than those correlations induced by the phase-space architecture itself.

4. The natural variables for the extended phase space are $\eta_i$, $\phi_i$, and $\log p_{t_i}$, where the index $i$ runs over the pieces of extended phase space. Note that these go directly over to essentially the *same* variables in a space-time description. This should be obvious for $\eta$ and $\phi$. And the variable "conjugate" to $\log p_t$ is $\log t_\perp$, where $t_\perp$ is the time it takes for the jet to evolve in a reference frame where the nearest neighboring jets emerge at right angles to the jet in question.

5. If hadronization of each color antenna or color-string is done in a collinear reference frame for the source-partons in a sensible way (e.g. azimuthal symmetry is enforced), the peculiar QCD correlations such as the "string effect" will be enforced.

It is my opinion that these are the essential features necessary and to a great extent sufficient for describing multiparticle production successfully from the QCD point of view. A detailed model which respects these features I would tend to trust; one which does not I would question.

## 3. Space-Time Geometry and HBT

An important goal of multiparticle dynamics is to delineate the space-time evolution of the production process. In heavy-ion collisions such a description is vital. Since even a proton has at least 3 constituent quarks, the importance of geometry in the (transverse) impact plane is likewise clearly important—more important I think than is, on average, recognized. And even in $e^+e^-$ annihilation the space-time evolution is decidedly nontrivial.

In the perturbative domain the evolution of an initially small system will be near the light-cone, because the relevant constituent partons can be regarded as massless: everything goes out with the speed of light from an initially compact source. If, as is generally assumed, the transition from parton system to produced-hadron system occurs quickly in its local rest frame, newly formed hadrons must likewise be found near the light-cone.

This does not mean that the laboratory time-scales are short. Many of these subsystems are extreme-relativistic, and they can evolve over large distances. This implies that at a given time the geography of the intermediate system is an expanding shell, with very little entropy in the interior. The surface of this shell is not simple in structure; in perturbative QCD it is most appropriately described in terms of the extended phase-space mentioned in the previous section. For multijet configurations there will be a variety of time scales for hadronization associated with the intrinsic relative $p_t$'s of the subsystems.

A very powerful tool for investigating space-time evolution is that of Bose-Einstein correlations. This was admirably reviewed by Eddi de Wolf[10] in this symposium. But both he and I however share a concern that the space-time dependence of popular source-distribution for hadron-hadron collisions (*not* ion-ion) may not be very realistic. Static gaussian or longitudinally expanding gaussian distributions seem to be the order of the day. Hollow shells would at the least imply Bessel-functions appearing in the HBT analysis. But I look in vain to find them.

As described so well by Seyboth,[11] the HBT analyses used in heavy-ion collisions are quite sophisticated—to me they appear to be in a more advanced state than their counterparts in hadron-hadron collisions. In present-day ion-ion collisions, some effects of transverse expansion are included, but the longitudinal expansion effects are dominant.[12] There are no hollow shells needed and gaussians abound.

One may question whether even for hadron-hadron collisions there really is a large transverse distance scale for hadronization. In the neighborhood of a jet the answer must be yes; the hadronization time in order of magnitude is one fermi per GeV of $p_t$. And even in jetless final states the radius can be large if the local multiplicity density is large. A reasonable picture of the purported expanding shell,[13] within some solid angle $\Delta\Omega$, as seen just after hadronization and taken (without loss of generality) to be at 90° emission angle, is a single layer of closely packed pions. This gives for the multiplicity $\Delta N$ in that solid angle

$$\Delta N = \left(\frac{dN}{d\eta\,d\phi}\right)\Delta\Omega = \frac{R^2 \Delta\Omega}{\pi r_\pi^2}\ . \qquad (2)$$

In other words

$$(R^2)_{\text{hadronization}} \approx (1.2f)^2\,\frac{dN}{d\eta\,d\phi} \qquad (3)$$

which is not terribly large on average. This formula, by the way, is interesting in its own right and is not well tested by existing data.

The thickness of the shell could be as small as $0.2 - 0.4f$ if the collisions occurred between only one pair of constituent quarks; a more typical estimate might be $0.5 - 1.5f$. But even if the mean radii are moderate, the fluctuations to large radii may be significant.

In any case, the HBT analyses do remain very important, especially at collider energies. Basic recommendations include

1. Always set up the HBT theoretical analysis in a frame where the pair which is to be observed emerges at 90° to the beam direction or thrust axis. This best decouples geometry effects from the intrinsic physics, and no generality is lost.

2. Subdivide the data with respect to the magnitude of associated multiplicity, and measure the correlations as function of the local $dN/d\eta \, d\phi$.

3. Also look at the $p_T$ dependence of the effects.

4. Take expanding-shell geometry for the source seriously. The formalism does exist[14] but to my knowledge has not been used very much.

5. In general, capitalize on the analysis methodology now extant in the heavy-ion community.

## 4. Multiplicities, Intermittency, Correlations

The question of intermittency and multiparticle correlations has been a dominant one for several years ever since the pioneering work of Bialas and Peschanski.[15] At this meeting, my (limited) perception of the situation goes as follows:

1. In ion-ion collisions, the only important correlation structure seen is intrinsically 2-body, short range in rapidity. As mentioned before, the HBT analyses are refined and the results intelligible.

2. Correlation structures seen in $e^+e^-$ are "understood" in the sense that they can be reproduced by Monte Carlo with reasonable (mostly perturbative or quasi-perturbative QCD) inputs. A possible exception is HBT phenomena, which are hard to simulate via Monte Carlo. But even these seem under reasonable control.

3. In hadron-hadron collisions the situation is less clear. Here the correlation plots, *e.g.* $\log F_2$ versus log of the bin size, shows smooth behavior across a wide range of scales, from the 1 GeV perturbative-QCD scale down to the 30-50 MeV scale of HBT Bose-interference. Good 3-dimensional analyses with momentum measurement exist.[16] The correlation is strongest amongst like-charge systems, indicating the importance of Bose-Einstein correlations.

The smooth behavior of these correlations over such a wide range of scales has produced a multitude of analyses using the notions of fractals, multifractals, and self-similar behavior. It is unclear how relevant these are, given the existence of all kinds of scales between 30 GeV and 1 TeV, in particular the 140 MeV pion-mass scale and the 300-500 MeV confinement scale. However, the data do exhibit the regularity and, as discussed by Bialas in his fine review,[17] there remains the issue of whether the regularity invites a simple explanation or is just an accident.

Bialas' search for simplicity leads him toward a space-time description, with the source of self-similar behavior being power-law fluctuations in the hadronization radius. Arguably this parameter, influenced by the dynamics at all scales, might be relatively immune to scale dependent effects.

In terms of my favorite expanding-shell picture, low-mass pions which possess the Bose-Einstein enhancement have to be produced at large formation radii for the geometrical reasons described in the previous section. These large radii are in turn most easily created by underlying minijets, thereby establishing a possible connection between the low pair-mass scale and the short-distance scale of QCD minijets. A test of this might lie in the dependence of the correlation radius, for fixed dipion mass and pair orientation, upon pair transverse momentum. So again it seems to me that for this problem an improvement in sophistication of the HBT technology along the lines already mentioned might also be of use.

Amongst the specific descriptions of the multiparticle correlation data, the Monte Carlo simulations lead the way. But there are many, including myself, who yearn for analytic methods that reveal the essence of the phenomena. So there exist a variety of cascade models, many of which carry the partonic branching processes to very low mass scales and then invoke "local parton-hadron duality." Of these I have mixed feelings. My reservations can be most easily expressed by consideration of the simulation of the process $e^+e^- \to$ hadrons at cms energies from 500 MeV to 5 GeV. It should be relatively easy to create a parton-cascade model that fits the multiplicity data: at low energies $e^+e^- \to \rho \to \pi^+\pi^-$ on the one hand, and $e^+e^- \to q\bar{q}$ on the other. At the high mass end, one parameter at most is needed to set the parton multiplicity relative to pion multiplicity correctly. In between everything should be all right because in each case the hadron/parton production is just phase-space—jet structure is only barely perceptible at cms energies of 5 to 6 GeV.

But while the parton-cascade picture is sufficient to generate the right correlation structure, it is not at all obvious to me it is necessary, nor that it faithfully images the detailed dynamics.

## 5. Disoriented Chiral Condensate

There were no detailed presentations at the Symposium of the status of theory and/or experiment regarding the search for disoriented chiral condensate (DCC). This happens to dominate my own interests these days and there actually has been an exponential growth of interest in this subject in the last two years. However, for lack of space I will not attempt much of a review here; recent summaries do exist elsewhere.[18,19]

DCC is a conjectured region of space-time containing strong-interaction vacuum but with a non-standard chiral orientation. In the language of the linear $\sigma$-model the order parameter

$$\langle \Phi \rangle = \langle \sigma + i \vec{\tau} \cdot \vec{\pi} \rangle \tag{4}$$

which usually point in the $\sigma$ direction is presumed to be in some other direction. At the most primitive, naive level, all the vacuum inside the aforementioned expanding shell might at early times have this property. (This is the "Baked Alaska" scenario.[20]) When the hot shell hadronizes and breaks up, the disorientation is radiated away in Goldstone modes (pions) of a common (cartesian) isospin. Such coherent pulses of semiclassical pion field would lead to anomalously large fluctuations event-to-event in the ratio of neutral to charged pions produced. Specifically with the definition of neutral fraction:

$$f = \frac{N_{\pi^0}}{N} \tag{5}$$

and with

$$N = N_{\pi^0} + N_{\pi^+} + N_{\pi^-} \tag{6}$$

the distribution for fixed $N$, however large, is

$$\frac{1}{n}\frac{dn}{df} = \frac{1}{2\sqrt{f}} \tag{7}$$

very different from the binomial distribution given by common sense (and for that matter by all simulations).

Quite a lot has been accomplished in developing these ideas in the last two years, although most of the original ideas, including the inverse square root formula, go back much further.[21,22] A brief summary (for hadron-hadron collisions; there is also work on ion-ion collisions) is as follows:

1. At an appropriate early time in the collision the energy density, assumed to be originally very high, decreases to values appropriate to the chiral phase transition.

2. Thereafter the collective variables $\sigma$, $\vec{\pi}$ are used to track the evolution of the chiral field.

3. While the chiral fields fluctuate a great deal, on average they initially have small vacuum expectation values; they lie "on top of the Mexican-hat potential."

4. Because of this the squared masses of the chiral fields are temporarily negative and there is exponential growth, especially at longer wavelengths. This evolution is tracked with the linear sigma-model from short proper times (*i.e.* near the light-cone) inward.[23]

5. When the field has "rolled" into or near the minimum of the potential well, the nonlinear sigma-model may be used. A class of simple solutions ("Anselm class"[22]) exist in this limit, which help to define and describe the DCC evolution in Minkowski space-time.

6. As proper time increases further, pion mass effects become important. At this point one can match the fields to the final state of a gas of emitted, freely propagating physical pions.

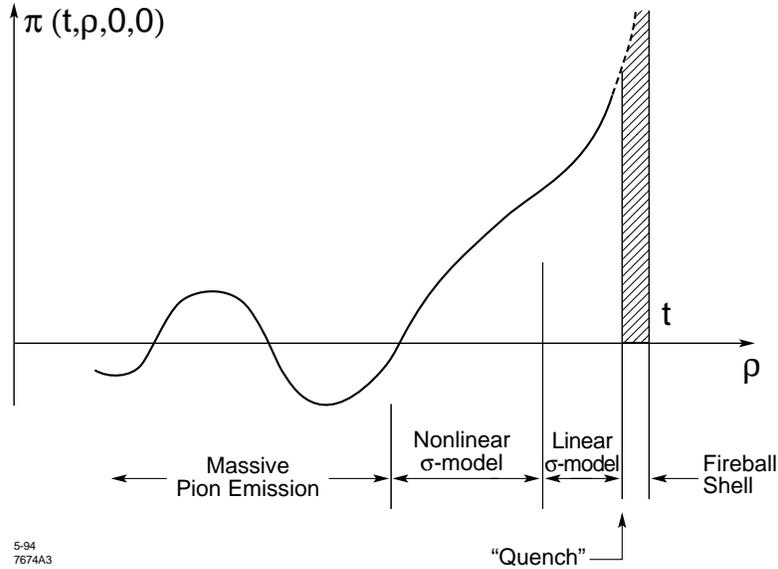

Figure 2. Structure of DCC "wake-field" following behind a pulse of partons produced at 90° to the beam direction. The dynamical descriptions believed appropriate are also indicated.

The DCC effects are biggest when the expanding system is out of equilibrium ("quench" scenario[24,25]). In space-time all the action is near the light cone, and so the volume in which DCC can be expected is limited to a thickness of $\lesssim 1-2$ $f$ in proper time units and a solid angle (at 90° production angle) of $\lesssim 1-2$ steradians. Nevertheless this space-time region may be big enough to make the effect observable. The DCC might be viewed as a (coherent) "wake-field" on the inner side of the expanding shell of ordinary produced partons, which eventually detaches inward (relative to the lightfront) as semirelativistic or nonrelativistic (at 90° emission) groups of pions emitted into the interior of the light-cone (Fig. 2).

This is an extremely speculative topic. It cannot live or die on theory alone; help from experiment is essential. It is my understanding that the heavy-ion community intends to undertake a DCC search in upcoming CERN runs, and we (test/experiment T864) are undertaking a search in the Fermilab collider,[26] in a phase-space region where cosmic-ray data hint at an effect.

## 6. Deep-Inelastic Scattering at HERA

Many interesting results from the HERA program were presented, and again there is not enough space here to make a serious summary. Much of the interest is driven by the rather dramatic rise in $F_2$ with $W^2$, the squared mass of the final state hadronic system. While this rising behavior was anticipated by theory—it clearly has to do with production of gluon jets into the final state phase space—the questions which are raised seem to push beyond the boundaries of theoretical control.

Among these questions are the following:

1. $F_2$ at fixed $Q^2$ seems to be rising as a power of $x^{-1}$, or of $W^2$, with an exponent somewhere between 1.25 and 1.5. This behavior cannot continue indefinitely; for a given $Q^2$ when does the growth saturate?? A related way of expressing this is as follows: where is the boundary (in $Q^2$) between the soft-Pomeron energy-dependence seen in photoproduction (exponent $\lesssim 1.08$) and the hard-Pomeron dependence (exponent $\sim 1.4$) apparently seen at large $Q^2$?

2. What is the detailed nature of the hadronic final states that leads to this strong energy dependence?

3. What is the role of the diffractive contribution and how does it evolve?

I will comment on these questions in reverse order, because it is really only the first which will be discussed in any detail at all:

3. Diffractive final states, by definition, are those containing rapidity gaps. Important discriminants of the dynamical mechanisms are[27]

   *a*) whether there are leading dijets in the photon fragmentation region, and

   *b*) whether the typical dijet mass is comparable to or less than $Q^2$, or whether it is typically large compared to $Q^2$.

However, there is no time here to delve into the theoretical options available in interpreting the observed hard diffraction. It is in my opinion a subtle, but most significant phenomenon well worth the trouble needed to properly understand it.

2. The nature of generic multiparticle final states in deep inelastic scattering is, at small $x$, expected to be a hybrid of $e^+e^-$ final-state properties and generic hadron-hadron final states. The general expectation,[28] in the absence of extra QCD jets, is that the region of the lego plot near the photon fragmentation region behaves as in $e^+e^-$ annihilation, while the rest behaves as in $h-h$ collisions. The dividing line in the lego plot, in any frame of reference for which virtual photon and proton are collinear, is a distance $\sim \ell n\, Q^2$ from the photon leading fragments and a distance $\ell n\, 1/x$ from the proton leading fragments. (Note that the sum is $\ell n\, W^2$ as required.) In the HERA laboratory frame, half of the $e^+e^-$-like phase space appears in the "struck quark jet," and the rest in a rapidity interval of width $\approx \ell n\, Q$ adjacent to the quark jet.

It would be very interesting to do the classic multiparticle production analyses (multiplicity distributions, factorial moments, clans, voids, etc.) on either side of the boundary and search for evidence for the expected transition from $e^+e^-$-like behavior to hadron-hadron-like behavior. In particular the void/gap analyses of Giovannini and co-workers[29,30] exhibited a strong difference in the two cases, suggesting that this will show itself clearly in the data. Unfortunately, for most of the existing HERA data, the boundary typically occurs at a small angle (large rapidity) relative to the proton direction. So in most of the accessible phase space, the analysis method should naturally imitate that in $e^+e^-$ annihilation. This seems to be what is happening.[31]

For the boundary to occur at 90° ($\eta = 0$), one should have $\nu = q \cdot p \cong 800\ GeV$, independent of $Q^2$. For $Q^2 = 30\ GeV^2$, this is $x \cong 0.02$, and would appear to be feasible to measure. (Note that these values are also accessible to the muon-scattering experiment E665 at Fermilab.)

*6.1 The rise of $F_2$: when does it stop?*

Here we ask the question: for fixed $Q^2$, at what value of $W$ does the strong power-law growth saturate? The inputs are simple, namely an assumed power-law growth, unitarity, and common sense. More specifically,

1. We assume that unless otherwise constrained $F_2 \sim W^\alpha$, with $\alpha$ chosen here to be 1. (This choice is not crucial; we really mean a value $\alpha$ in the range 0.5–1.) And by "unless otherwise constrained," we mean that this behavior cannot be true at large $x$, nor at very small $Q^2$, but only within some midrange in between. This region must of course be defined; this is the point of the exercise.

2. We ignore logarithmic dependence; it is only the principal power-law behavior (Regge-intercept) we are addressing.

3. We impose unitarity in a very simple way, by recalling an old observation of Gribov[32] based on generalized vector dominance. One may write

$$\sigma_T(Q^2, W^2) = P_{\text{had}} \cdot \sigma_{\text{had}} \qquad (8)$$

where

$$P_{\text{had}} = \frac{\alpha}{3\pi} \int_0^{\mathcal{O}(W^2)} \frac{dm^2 m^2 R(m^2)}{(Q^2 + m^2)^2} \equiv \frac{\alpha}{3\pi} \overline{R} \log \frac{W^2}{Q^2} \qquad (9)$$

is the probability that on arrival at the nucleon in, say, the fixed-target reference frame, the photon has fluctuated into a hadronic system. Note that this is essentially $(1 - Z_3)$: $R$ is the familiar sum over squared quark charges that defines the $e^+e^-$ cross-section to hadrons.

The quantity $\sigma_{\text{had}}$ is simply the absorption cross-section of this hadronic system on a proton. It is bounded by the geometrical cross section[*]

$$\sigma_{\text{had}} \lesssim \pi(r_\gamma^2 + r_p^2) \approx \pi r_p^2 = \frac{\sigma_{pp}}{2} \qquad (10)$$

where we use gaussian density distributions and assume the virtual-photon system is smaller than a proton. We now put in numbers, approximating $\log W^2/Q^2$ by a constant. Our fit (Fig. 4) for $F_2$ is, in the range of $Q^2$ and $W^2$ appropriate for the power-law growth:

$$F_2 \approx 0.4 \left(1 + \frac{W}{100\,GeV}\right) \leq \frac{Q^2}{4\pi^2 \alpha} \cdot \left(\frac{\alpha}{3\pi} \overline{R} \log \frac{W^2}{Q^2}\right) \cdot \frac{\sigma_{pp}}{2} . \qquad (11)$$

Taking

$$\overline{R} = 4$$

$$\log \frac{W^2}{Q^2} = 15 \quad \text{(to a factor 2)} \qquad (12)$$

$$\sigma_{pp} = 60 \pm 20\,mb$$

---

[*] We ignore the factor $\sim (\text{const})/Q^2$ associated with the "aligned jet" picture, because this is appropriate for soft final states, not the multigluon-jet states that creates the BFKL structure. See Ref. 27 for some more discussion of this point.

gives for the unitarity limit boundary

$$\left(\frac{W}{100\ GeV}\right) \lesssim \left(\frac{Q^2}{0.4\ GeV^2}\right). \tag{13}$$

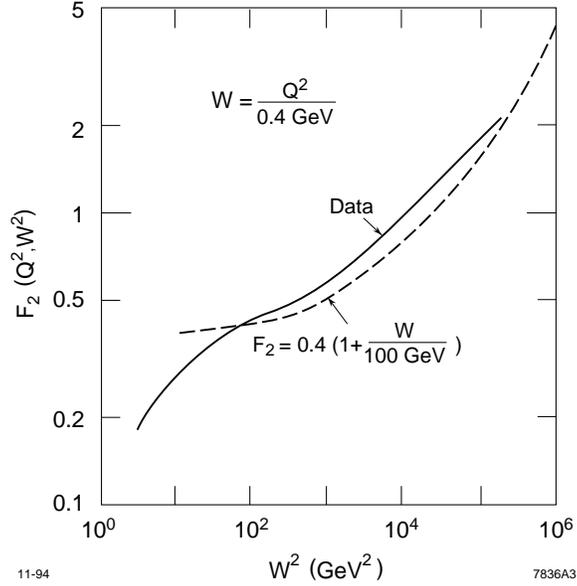

Figure 3. A sketch of the dependence of $F_2(W)$ with the constraint $W = (Q^2/0.4\,GeV)$. See the text for the motivation for this choice.

4. Now we apply common sense. In Fig. 4(a) is sketched a contour plot of $\ell n\, F_2$ in $\ell n\, Q^2 - \ell n\, W^2$ space under this assumption. To my eye it looks awkward, especially the rapid rise of $F_2$ with $Q^2$ for fixed $W$ in excess of 1 TeV. More reasonable to my eye is Fig. 4(b), based on the assumption that $F_2$ never exceeds 1% of the unitarity bound:

$$\left(\frac{W}{100\ GeV}\right) \lesssim 0.01 \left(\frac{Q^2}{0.4\ GeV}\right). \tag{14}$$

Of course this is only a matter of taste at this level. But with this choice there is nothing that occurs at higher energy other than an extension of what has been happening at accessible energies.

But irrespective of these details, the main issue is whether the unlimited growth of $F_2$ as $Q^2$ and $W^2$ mutually tend to infinity violates any fundamental principles. What is implied is that eventually an infinite number of gluons *per unit rapidity* are packed into a finite area in the transverse impact plane (remember, we are ignoring logarithmic growth of sizes here!), and we are assuming they do not recombine. This sounds hopelessly naive. But there is a chance it is not. After all the growth of $F_2$ is to be associated with production of gluons into the extended phase space, which with its anomalous dimension and fractal nature has infinite area per unit laboratory rapidity. No doubt the phase-space associated with *initial-state* partons must likewise be regarded as extended and fractal so it is possible that this infinite gluon population may still find an uncrowded home in its extended phase space, with no need for major amounts of recombination. But, at least for me, much more experience with this still somewhat unfamiliar language is needed before being fully comfortable with this conclusion.

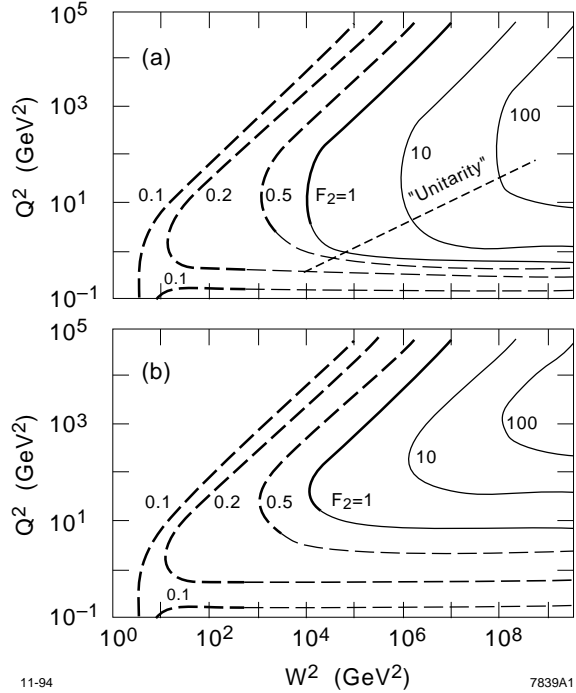

Figure 4. Contour plot of $\log F_2$ as function of $\log W^2$ and $\log Q^2$, assuming (a) the "unitarity" constraint, Eq. (13); (b) the "common-sense" constraint, Eq. (14).

I have been informed by Bo Andersson that the guess I have made regarding the shape and growth of $F_2(Q^2, W^2)$ is consistent with what he and his Lund colleagues are finding using the Lund version of extended phase-space. There were not many details in Andersson's report,[7] and I am not sure that I interpret correctly what they say. But maybe what is said is something like the following: The natural BFKL density of produced gluons in the lego plot exceeds what is naturally allowed from multi-Regge kinematics. The former is roughly determined by

$$\frac{dN_{\text{jets}}}{d\eta} \approx n \approx \frac{12\alpha_s}{\pi} \ln 2 \approx \frac{4.0}{\ln p_T^2} \qquad (15)$$

where $n$ is the BFKL exponent which purportedly controls the growth of $F_2$ with $W^2$.

On the other hand, in order to allow a straightforward application of multi-Regge kinematics, the spacing of gluons should be larger than $\ln p_T$

$$\Delta \eta \gtrsim \ln p_T = 0.5 \ln p_T^2 \ . \qquad (16)$$

However from Eq. (15) we see that the gluon spacing would like to be

$$\Delta \eta_{\text{Jets}} \approx 0.25 \ln p_T^2 \ . \qquad (17)$$

The $p_T$ behavior is the same, but the coefficient is too small by a factor two. This kinematic restriction is generic[33] and, according to Andersson, suppresses the size of the BFKL exponent to a value consistent with the soft-Pomeron energy dependence. This is a most interesting and important inference, which deserves much more critical attention.

## 7. Other Contributions.

In this summary talk I have omitted many interesting contributions. I cannot but help mentioning here a few, and apologize to those I have omitted. What follows is a series of brief advertisements, nothing more:

$p\bar{p} \to$ *forward $W+$ jet* (Geoffrey Forden): The DØ collaboration finds the jet is not dragged forward with the $W$ as much as perturbative QCD predicts; Andersson, in discussion comments, likes the result and stresses its importance.

*Heavy Ion Program* (Helmut Satz, Peter Seyboth): To me, impressible progress into a higher critical level, with more precise guidelines toward the identification of quark-gluon plasma in future experiments.

*Correlation analyses* (Wolfgang Ochs and Jacek Wosiek): New tools available for intermittency studies.

*Clans and rapidity gaps* (Alberto Giovannini, Sergio Lupia, Roberto Ugoccioni): Analysis of multiparticle spectra via clans provides a prediction of rapidity gap probability versus gap width. This is expected to be different in hadron-hadron collisions (Reggeized rho, omega exchange) and in $e^+e^-$ annihilation (pion form factor controls the dependence), and it is found to be so. I think this is a very nice result.

*Interpolation of correlation moments* (Rudy Hwa): An improved method to describe the low-multiplicity component of multiplicity distributions and moments.

*Jet quenching in ion-ion collisions* (Michael Plumer): A thorough analysis of a favorite subject of mine (that means I once worked on it myself).

*Entropy generation* (Hans-Thomas Elze): Not only was an interesting problem defined in this presentation—how the large final-state entropy of collision products evolves from the much smaller initial-state entropy—but impressive progress was made toward understanding it.

*Charm production dynamics* (Sergio Ratti): A rather high $p_t$ scale is observed in hadron-hadron collisions, especially in comparison with photoproduction.

*Strange baryon production at BNL* (Erik Gottschalk): Impressive results from a very high-statistics, high-rate experiment; there is likely to be much spectroscopy to be mined from this data set.

*Semileptonic B decays* (Leo Bellantoni): A very pretty analysis of ALEPH data.

*Deep Inelastic Muon Scattering* (Jorge Morfin, Jona Oberski): High quality data on shadowing, neutron/proton ratios and low $Q^2$ behavior from the Fermilab E665 and CERN NMC experiments provide important complementary information to what is being seen at HERA.

## Acknowledgments

It is on behalf of all participants that I thank Alberto Giovannini and his colleagues for their hard work in organizing so successfully this fine symposium.